\documentclass[conference]{IEEEtran}
   \usepackage[pdftex]{graphicx}
   \DeclareGraphicsExtensions{.pdf,.jpeg,.png,.JPG,.jpg}
\usepackage{marginnote}
\usepackage{amsmath}
\usepackage{xfrac}
\usepackage{float}
\usepackage{cite}

\newcommand{\sss}{\scriptscriptstyle}
\newcommand{\sst}{\scriptstyle}

\newcommand{\stext}[1]{\sss \text{#1} \sst}

\newcommand\numberthis{\addtocounter{equation}{1}\tag{\theequation}}

\usepackage{eso-pic}

\makeatother

\AddToShipoutPicture{
	\raisebox{2\baselineskip}{\makebox[\paperwidth]{\begin{minipage}[t][0.5cm][b]{21cm}\centering \scriptsize \copyright 2019 IEEE. Personal use of this material is permitted. Permission from IEEE must be obtained for all other uses, in any current or future media, including reprinting/republishing this material for advertising or promotional purposes, creating new collective works, for resale or redistribution to servers or lists, or reuse of any copyrighted component of this work in other works.\end{minipage}
}}}

\begin{document}
%
\title{Diffraction Gratings for Uniform Light Extraction from Light Guides}

\author{\IEEEauthorblockN{Micheal McLamb, Yanzeng Li, Serang Park, Marc Lata, and Tino Hofmann}
	\IEEEauthorblockA{Department of Physics and Optical Science,\\ University of North Carolina at Charlotte,\\ 9201 University City Blvd, Charlotte, NC, 28223\\
		Email: mmclam10@uncc.edu}}


%


\maketitle

\begin{abstract}
A theoretical approach for uniformly extracting light propagating through a light guide plate is developed here. Typically, liquid crystal display modalities utilize a backlit lighting system illuminated from the edge with light emitting diodes. The backlight acts as a light guide plate and is coupled with a diffuser sheet along the surface to uniformly extract the light. Our approach employs sub-micron diffraction gratings and eliminates the need for diffuser sheets, while ensuring the uniform extraction of light. The physical dimensions of the grating are varied along the surface of the light guide plate to control the diffraction efficiency, thus determining the local viewing angle and emitted intensity.
\end{abstract}


%
\IEEEpeerreviewmaketitle

\section{Introduction}

Backlights are the prevailing illumination sources for liquid crystal displays. They are found in televisions, laptops, and cellular phones, for instance, and are designed to provide a spatially uniform light source for the display. Previous generations of liquid crystal displays relied on cold cathode fluorescent backlighting . Modern LCD displays now employ LED backlighting, which provides substantial advantages over cold cathode fluorescent sources mainly due to lower power consumption and increased lifespan \cite{minano2005high}. In contrast to cold cathode fluorescent backlighting, LED based sources illuminate the display through light guide plates (LGP) from the sides. Thus, such display designs require large area diffractive optical components to uniformly distribute the light from the LED across the surface of the display. Diffuser sheets and prism films of varying transmission have been used for this purpose \cite{parikka2001deterministic}. While layered diffuser sheets are very simple to design they provide uniform illumination at the cost of additional thickness, increased power loss, and no control over the viewing angle. On the other hand, diffractive optical components offer an alternative approach for light extraction from the LGP at uniform intensity and viewing angle. Among numerous different grating geometries, which have been employed by various authors to attain uniform extraction, sub-micron diffraction gratings have been identified as the most promising technique \cite{Cornelissen2013diffraction,imai2008illumination,wang2015high}. 

Typically, the geometries of the diffraction gratings are optimized using various methods, such as the Taguchi method and rigorous coupled wave analysis, in simulation software until uniform extraction requirements are met \cite{park2013design,park2007grating, fang2014study}. In this paper an alternative theoretical approach is provided in order to develop diffractive optics that provide light extraction at uniform intensity and viewing angle. In contrast to widely applied techniques, the method demonstrated here is independent of the structural parameters of the grating. This is achieved by dividing the grating into areas for which the diffraction efficiency is uniform. By following discretization, a change in the structure height between different diffactive areas is sufficient in order to provide a  light extraction at a uniform angle and intensity across the length of the LGP. In addition to simplifying the numerical optimization for such a geometry, this design also reduces the fabrication complexity.      



\section{Theory}

\begin{figure}[bth]
\centering
\includegraphics[width=\columnwidth, trim=0 130 0 100,clip]{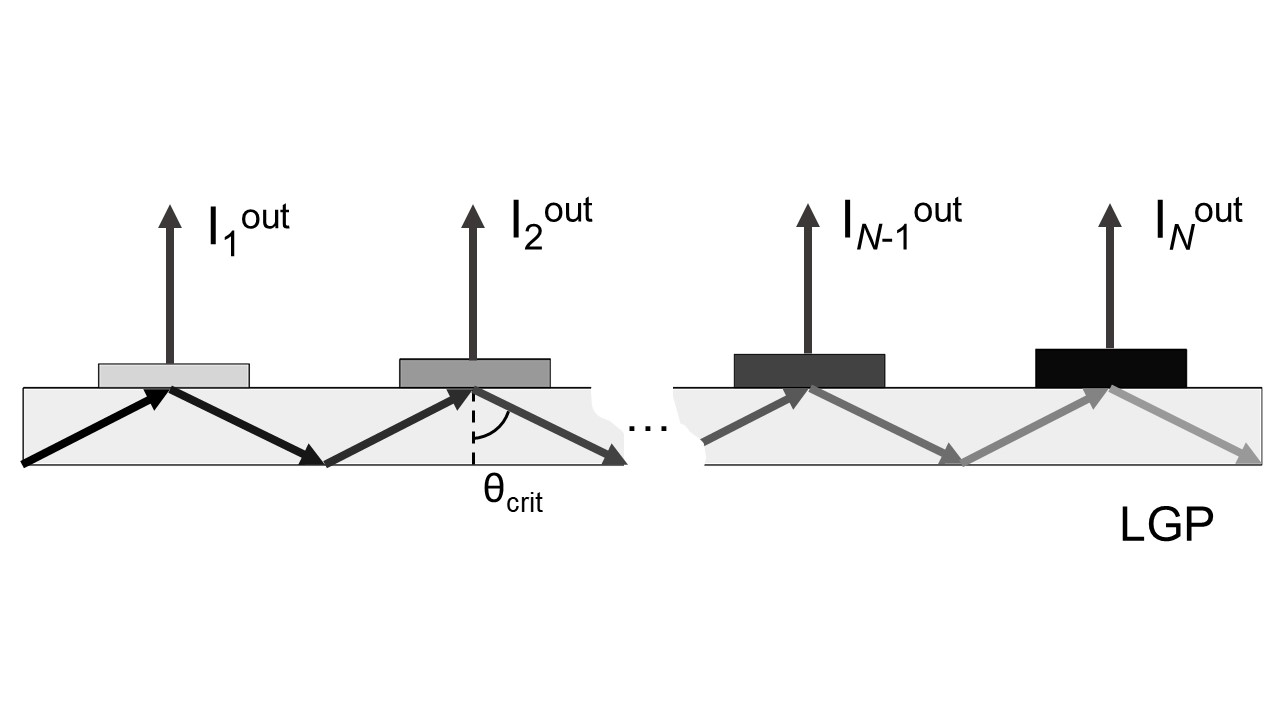}
\caption{Areas with diffractive gratings (shaded rectangles) are patterned along the surface of a light guide plate to enable uniform light extraction ($I^{\stext{out}}_i$~=~const.). The efficiency of the diffraction gratings used for the light extraction increases along the direction of propagation, represented as increased shading of rectangles, to produce uniform extracted angle and intensity $I^{\stext{out}}_1=I^{\stext{out}}_2=\ldots I^{\stext{out}}_N=$~const.}
\label{fig_setup7}
\end{figure}

The uniform light extractor design demonstrated here is a modified light guide plate as shown in Fig.~\ref{fig_setup7}. As the light propagates through the LGP it will interact with diffractive structures patterned along the surface indicated as differently shaded rectangles in Fig.~\ref{fig_setup7}. For each interaction, the light will be extracted according to the diffraction efficiency of the structures within the diffractive area \cite{parikka2001deterministic,pan2012light}. As the light is depleted by a grating module it becomes apparent that the light inside the LGP will begin to decrease. If the light extractor is to extract uniformly it becomes obvious that the diffraction efficiency of each module must increase to compensate for the remaining depleted light.\\

\begin{figure}[h]
\centering
\includegraphics[width=\columnwidth, trim=0 200 0 137,clip]{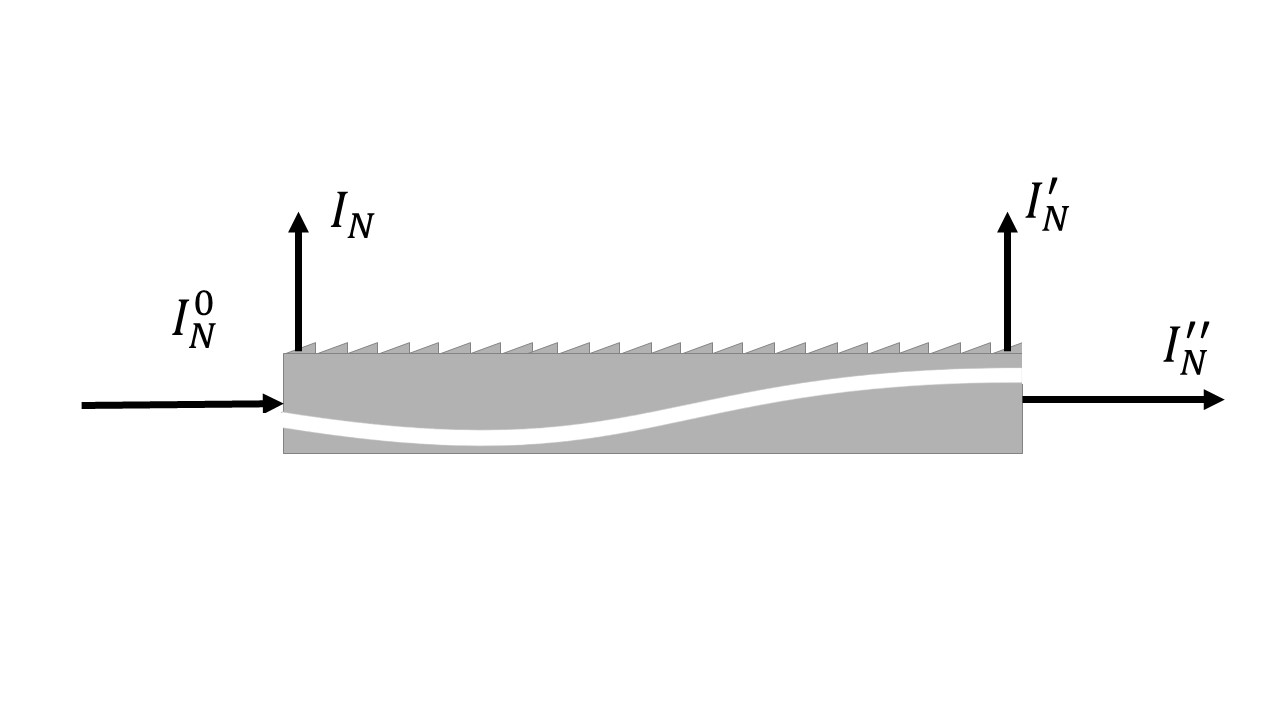}
\caption{Schematic illustrating the light intensity entering and exiting a an area which a homogeneous diffractive grating as shown in Fig.~\ref{fig_setup}. $I_N^0$ indicates incoming intensity, $I_N$ the diffracted intensity at first structure, $I_N^\prime$ the intensity at the end of the patch, $I_N^{\prime \prime}$ is the remaining light after traversing the diffractive area.}
\label{fig_setup}
\end{figure}

It can be seen from the schematic in Fig.~\ref{fig_setup} that the intensity at the final structure is related to the initial intensity diffracted at the first structure. As light propagates across a diffractive zone, an intensity drop will occur. The change in intensity for each patch will vary slightly according to the number of interactions between the beam and diffractive structure $b$ and the total number of patches the TIR light must propagate through,
where $a_N\%$ is the diffraction efficiency. The intensity for the final interaction with a diffractive structure can be written in terms of the diffraction efficiency as,

\begin{equation}
I_N^{\prime} =I_N^0a_N(1-a_N)^{b-1}.
\end{equation}

\noindent The intensity a patch will extract can be written as the sum of intensities that interact with the surface structures,

\begin{align*}
I_N^{\stext{out}} &= \sum_{i=1}^{i=b}I_N^0a_N(1-a_N)^{i-1},\\
I_N^{\stext{out}} &= I_N^0(1-a_N)^b.\label{eqn:sum} \numberthis
\end{align*}

\noindent An additional two constraints must be introduced here. Firstly, the total intensity emitted by each patch must be equal to ensure uniform light extraction,

\begin{equation}
I_{1}^{\stext{out}} = I_{2}^{\stext{out}} = ...=I_{N-1}^{\stext{out}} = I_{N}^{\stext{out}}. \label{eqn:eql}
\end{equation}

\noindent Secondly, as totally internally reflected light is extracted by the diffractive elements it becomes clear that $I_N^0$ decreases  as it propagates away from the source. $I_N^0$ is then the total intensity introduced into the LGP excluding the output intensities extracted by each patch,

\begin{equation}
I_N^0 = I_1^0 - I_2^{\stext{out}} - I_3^{\stext{out}} - ...I_{N-1}^{\stext{out}}.\label{eqn:In_diff}
\end{equation} 

\noindent Note that each output intensity is equal due to the constraint placed by Eqn.~(\ref{eqn:eql}),

\begin{align*}
I_N^0 &= I_1^0 - (N-1)I_1^{\stext{out}},\\
I_N^0 &= I_1^0 - (N-1)\sum_{i=1}^{i=b}I_1^0a_1(1-a_1)^{i-1},\\
I_N^0 &= I_1^0(N-1)(1-(1-a_1)^b). \numberthis
\end{align*}

\noindent The diffraction efficiency of any patch and its dependence on the intital patch transmission can be expressed as a function of the patch number using Eqns. (\ref{eqn:sum}), (\ref{eqn:eql}), and (\ref{eqn:In_diff}).

\begin{gather*}
a_N = 1-\Big[1-\frac{1-(1-a_1)^b}{1-(N-1)(1-(1-a_1)^b}\Big]^\frac{1}{b}. \numberthis
\end{gather*}

\noindent It must be noted that these expressions were derived under the assumption that only the 1st order transmission is extracted and all other higher diffractive orders are suppressed. This can be achieved by selecting a suitable geometry.

\begin{figure}[h]
\centering
\includegraphics[width=\columnwidth, trim=0 114 0 151,clip]{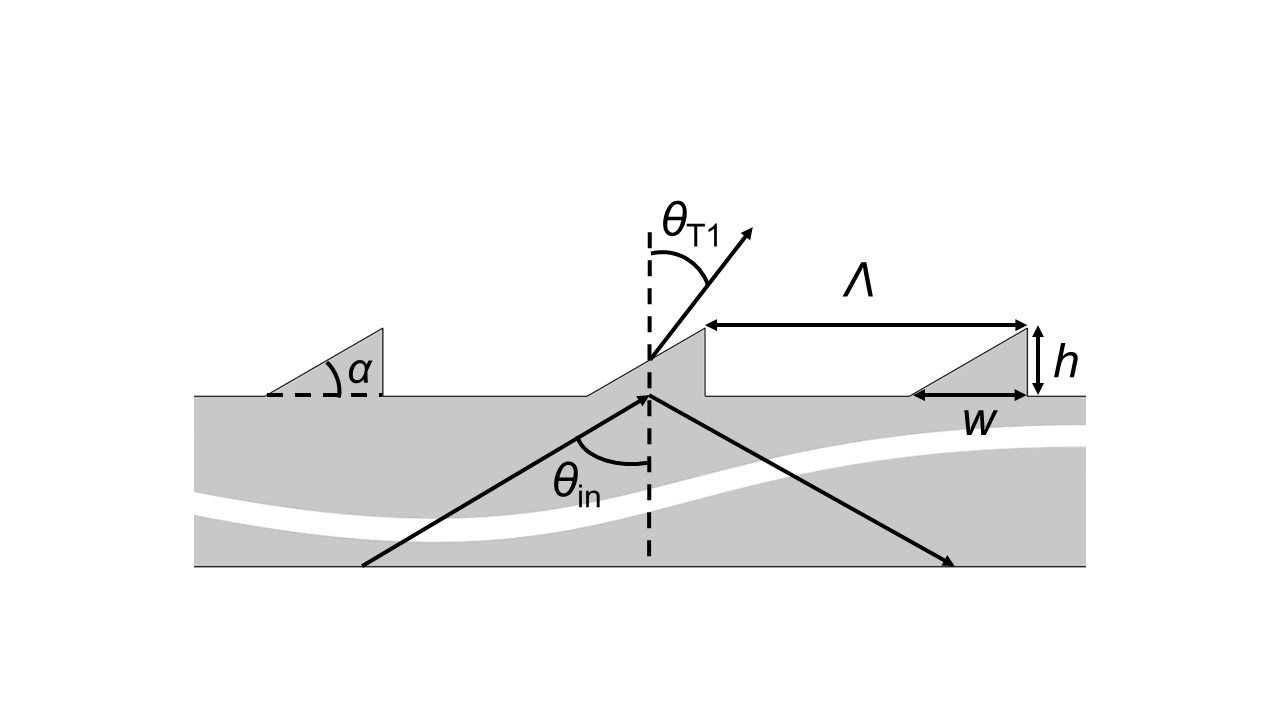}
\caption{Geometric parameters of the blazed diffraction gratings used for light extraction. While blaze angle $\alpha$, structure legnth $w$, structure height $h$, and periodicity $\Lambda$ can be used to change the diffraction efficiency of the gratings, only $\alpha$, is used here. Note, that higher diffraction orders are suppressed for gratings with periodicity less than the wavelength.}
\label{fig_geometry}
\end{figure}

A blazed angle geometry was selected for the arbitrary structures shown in Fig.~\ref{fig_setup}. The periodicity can be decreased below the incoming light wavelength to suppress higher orders. A blazed angle grating further suppresses the $-1^{ \stext{st}}$ order reflection while maintaining the $1^{ \stext{st}}$ order transmission due to the slanted structure of the blaze grating and orientation to the incoming light \cite{levola2007replicated}.  The viewing angle is determined using the grating equation for transmission,

\begin{equation}
n\sin(\theta_{\stext{in}}) + \sin{\theta_m} = \frac{m\lambda}{\Lambda}.
\end{equation}

\noindent When adjusting the diffraction efficiency for the proceeding patches only the blaze angle $\alpha$ should be altered to avoid changing the viewing angle \cite{parikka2001deterministic}.

\section{Numerical Simulation}

The geometric parameters of a blazed structure can be optimized using finite numerical simulations. Consider extraction from a glass substrate with refractive index $n$ equal to 1.5 at 450 nm. Suppose it is desired to extract 30\% of the introduced light from the patch. This can be achieved by placing 3 diffractive patches in a linear fashion along the surface of the glass substrate. Each patch will extract 10\% of the total intensity inside the LGP. That is to say that $I_1^{\stext{out}} = 0.1I_1^0$ . Using equation (2), the diffraction efficiency of the initial patch's structures can be determined,

\begin{equation}
a_1 = 1-(1-\frac{I_1^{ \stext{out}}}{I_1^0})^{\frac{1}{b}}.
\end{equation}

\noindent The design parameter $b$ represents the number of interactions a beam at a certain angle of incidence will have with the surface paired with the diffractive structures. Increasing $b$ results in a shorter patch and higher diffraction efficiency required for uniform extracted intensity. For this particular example, $b = 5$. Using these parameters $a_1 = 0.0209$. A simple blazed structure unit cell is then constrcuted in COMSOL Multiphysics for numerical simulation. Structures are scaled to the order of the wavelength of incident light. The source is monochromatic at $\lambda = 450 $ nm. The structure's periodicity is reduced until diffractive orders greater than 1 and less than -1  are suppressed. The blaze angle is adjusted to control and reduce back reflections due to the $-1^{\stext{st}}$ reflection. \\

\begin{figure}[h!]
\centering
\includegraphics[width=\columnwidth, trim={130 25 75 60},clip]{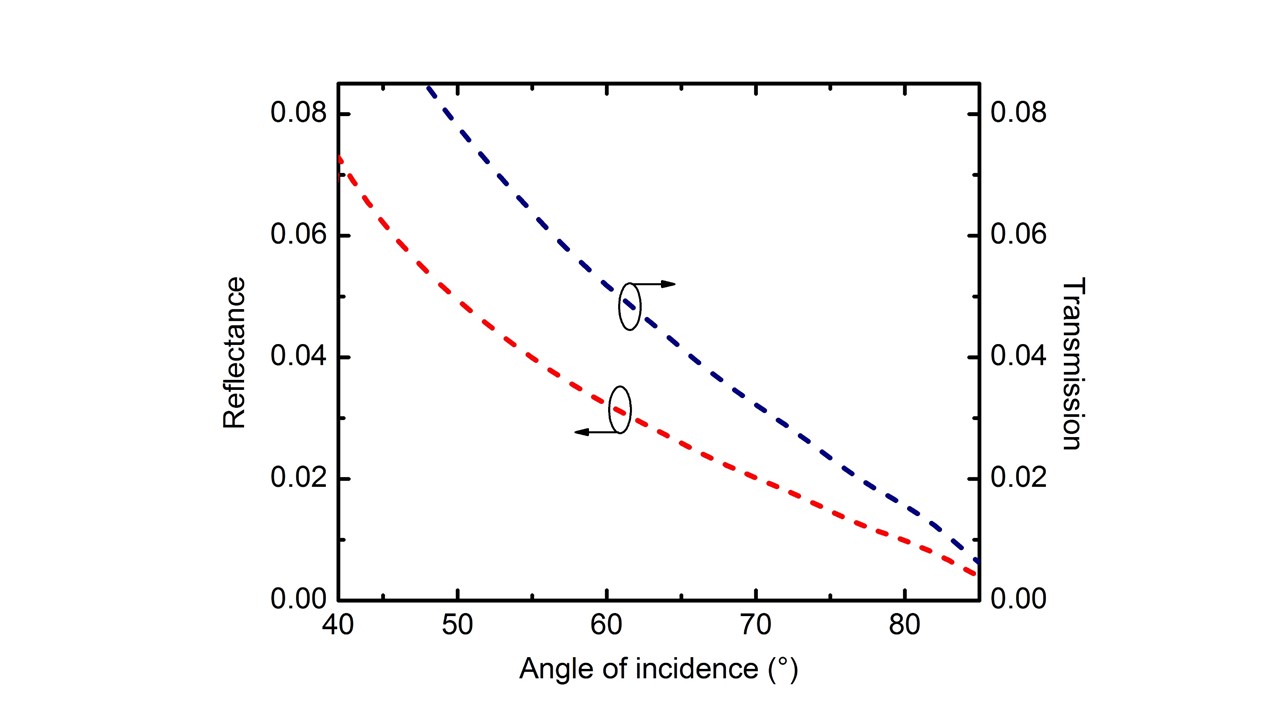}
\caption{Reflection and transmission of  30$^{\circ}$ blaze angle geometry unit cell. All negligible higher orders were suppressed for brevity. There is approximately 2\% transmission at a 75$^{\circ}$ angle of incidence.}
\label{fig_sim}
\end{figure}

\section{Results and Discussion}

The results for a blazed geometry confirm that the desired diffraction efficiency of the first patch can easily be met at a $75^\circ$ angle of incidence. To acheive this diffraction efficiency the blazed structure should have geometric parameters $\alpha = 30^\circ$,  $w =173$ nm, $h = 100 $ nm, and $\Lambda = 300 $ nm. This process can be repeated for the proceeding patches. To  alter the diffraction efficiency for the proceeding patch only the blaze angle should be adjusted to avoid altering the viewing angle. The length of the patches can now be calculated using the number of interactions $b$, the height of the substrate, and the angle of incidence. For a substrate of $y = 0.5 $ mm the length a beam will travel $x$ before it interacts with a surface can be expressed using trigonometry, $x = y\tan(\theta_i)$, $x = (0.5$ mm$)\tan(76^\circ) = 2$ mm, $l = 2x(b-1)$, $l = (2)(2$ mm$)(5-1)$, $l = 16 $ mm. Each patch must contain structures of a constant diffraction efficiency per patch as prescribed by equation (8), arranged in lengths of 16 mm. If three of these patches are placed in the direction of propagation, 30\% of the introduced light will be extracted.

\section{Conclusion}

It has been shown that diffraction efficiency requirements for light extraction at uniform intensity can be independently derived of the geometry to be used. The geometry must be designed and optmized according to the assumption that higher diffractive orders are negligible, but beyond this, parameters can be freely altered to adjust diffraction efficiency as needed. The $-1^{\stext{st}}$ reflection can be further suppressed by incorporating some slanted geometry such that it becomes negligible.
 This method allows the user to control intensity extraction and viewing angle without the deposition of films or extensive tuning through simulation and ultimately saves time by eliminating unnecessary large scale simulations involving critical features. Once a required diffraction efficiency is determined and reinforced with a suitable geometry only adjustment of a single geometric parameter, in this case the geometry's height, is needed to fulfill subsequent transmission requirements. For optimization of these gratings, only simulations of a single unit cell will suffice for each patch. The intensity drop across a particular patch can be mitigated through prudent selections in the design parameters, namely, the number of interactions a particular mode will have with surface structures and the total intensity extracted by each patch. The intensity drop can be sharply reduced to be negligible. In summary, light extraction at uniform intensity and viewing angle can be achieved by determining the initial diffraction efficiency independently of geometry. This diffraction efficiency then becomes the target parameter for designing and optimizing a geometry for the grating.

\section*{Acknowledgment}
	The authors would like to thank Steve Rosenblum and Indrani Bhattacharyya, from within the Center for Metamaterials, for their valuable comments and suggestions. The authors are grateful for support from the National Science Foundation (1624572) within the I/UCRC Center for Metamaterials, the Swedish Agency for Innovation Systems (2014-04712), and the Department of Physics and Optical Science of the University of North Carolina at Charlotte.



%


\end{document}